\documentstyle[12pt]{article}
\begin{document}
\setcounter{page}{1}
\begin{center}
{\bf Kramers-Wannier symmetry and strong-weak-coupling duality
in the two-dimensional $\Phi^{4}$ field model}
\bigskip

{\bf B.N.Shalaev}

\bigskip
{\it A.F.Ioffe Physical\& Technical Institute,
Russian Academy of Sciences,
Polytechnicheskaya str. 26,  194021 St.Petersburg (Russia),

Istituto Nazionale di Fisica Nucleare, Struttura di Pavia,
Via Bassi,6-27100, Pavia (Italy)

and Max-Planck Intsitute, Heisenbergstrasse,1, D-70529, Stuttgart,
(Germany)}
\end{center}

\begin{abstract}
It is found that the exact beta-function $\beta(g)$ of the continuous
2D $g\Phi^{4}$ model possesses two types of dual symmetries,
these being the Kramers-Wannier (KW) duality symmetry and
the weak-strong-coupling symmetry $f(g)$, or S-duality.
All these transformations are explicitly constructed.
The $S$-duality transformation $f(g)$ is shown to connect domains
of weak and strong couplings, i.e. above and below $g^{*}$ with $g^{*}$
being a fixed point.
Basically it means that there is a tempting possibility to compute
multiloop Feynman diagrams for the $\beta$-function using high-temperature
lattice expansions. The regular scheme developed is found
to be strongly unstable. Approximate values of the renormalized
coupling constant $g^{*}$ found from duality symmetry
equations are in good agreement with available numerical results.
\end{abstract}

\newpage
\section{\normalsize \bf INTRODUCTION }
\renewcommand{\theequation}{I.\arabic{equation}}
\setcounter{equation}{0}

The 2D Ising model and some other lattice spin models are known to possess
the remarkable Kramers-Wannier(KW) duality symmetry, playing an important
role in statistical mechanics, quantum field theory \cite{ab1,ab2,ab3}
as well as in the superstring
theory \cite{ab3a}. The self-duality of the isotropic 2D Ising model means
that there exists an exact mapping between the high-T and low-T expansions
of the partition function \cite{ab3}. In the transfer-matrix language this
implies that the transfer-matrix of the model under discussion is covariant
under the duality transformation. If we assume that the critical point is
unique, the KW self-duality would yield the exact Curie temperature of the
model. This holds for a large set of lattice spin models including systems
with quenched disorder (for a review see \cite {ab3,ab4}).
Recently, the Kramers-Wannier duality symmetry was extended to
the continuous 2D $g\Phi^{4}$ model \cite{ab5c,ab5d} in the strong-coupling
regime, i.e. for $g>g_{*}$.

This beta-function $\beta(g)$ is to date known
only in the five-loop approximation within the framework of conventional
perturbation theory at fixed dimension $d=2$ \cite{ab5,ab5b}.

The strong coupling expansion
for the calculation of the beta-function of the 2D scalar $g\Phi^{4}$ theory
as an alternative approach to the standard perturbation theory was recently
developed in \cite{ab5c,ab5d}.

It is well known from quantum field theory and statistical mechanics
that any strong
coupling expansions are closely connected with the high-temperature
(HT) series expansions for lattice models.
From the field-theoretical point of view the HT series
are nothing but strong coupling expansions for field models,
lattices to be considered as a technical device to define cut-off field
theories (see \cite{ab5c,ab5b} and references therein).

Calculations of beta-functions are
of great interest in statistical mechanics and quantum field theory. The
beta-function contains the essential information on the renormalized coupling
constant $g^{*}$, this being important for constructing
the equation of state of the 2D Ising model.
Duality is known to impose some important constraints on
the exact beta-function \cite{ab7b}.

In this paper we study other duality symmetries of the
beta-function $\beta(g)$ for the 2D $g\Phi^{4}$ theory
regarded as a non-integrable continuum limit of the exactly
solvable 2D Ising model. The main purpose is to construct exlicitly
the weak-strong (WS) coupling duality transformation $f(g)$
connecting domains of weak and strong couplings, i.e. above and
below $g_{*}$. The last transformation allows one to compute
unknown yet multiloop orders (6,7,...)
of the $\beta$-functions on the basis of lattice expansions
(\cite{ab5c,ab5d}).

The paper is organized as follows. In Sect. II we set up basic
notations and
define both the correlation length and beta-function $\beta(g)$.
In Sect.III  the duality symmetry transformation $\tilde{g}=d(g)$ is
derived. Then it is proved that $\beta(d(g))=d^{\prime}(g)\beta(g)$.
An approximate expression for $d(g)$ is also found.
Sect. IV contains an explicit derivation of the weak-strong coupling
transformation whilst in Sect.V in order to illustrate our approach
the sixth-order term of $\beta(g)$ is approximately computed.
Sec. VI. contains disussion and some concluding remarks.

\section{\normalsize \bf CORRELATION LENGTH AND COUPLING CONSTANT}
\renewcommand{\theequation}{2.\arabic{equation}}
\setcounter{equation}{0}

We begin by considering the classical Hamiltonian of the 2D Ising model
(in the absence of an external magnetic field), defined on a square lattice
with periodic boundary conditions; as usual:

\begin{equation}
H=-J\sum_{<i,j>}\sigma_{i}\sigma_{j}
\label{b1}
\end{equation}

\noindent
where $<i,j>$ indicates that the summation is over all nearest-neighboring
sites; $\sigma_{i}=\pm1$ are spin variables and $J$ is a spin coupling.
The standard definition of the spin-pair correlation function reads:

\begin{equation}
G(R)=<\sigma_{\bf R}\sigma_{\bf 0}>
\label{b3}
\end{equation}

\noindent
where $<...>$ stands for a thermal average.

The statistical mechanics definition of the correlation length is given
by \cite {ab8}

\begin{equation}
\xi^{2}=\frac{d\ln G(p)}{dp^{2}}|_{p=0}
\label{b4}
\end{equation}

\noindent
The quantity $\xi^2$ is known to be conveniently expressed in terms of the
spherical moments of the spin correlation function itself, namely

\begin{equation}
\mu_{l}=\sum_{\bf R} (R/a)^{l}G({\bf R})
\label{b5}
\end{equation}

\noindent
with $a$ being some lattice spacing. It is easy to see that

\begin{equation}
\xi^{2}=\frac{\mu_{2}}{2d\mu_{0}}
\label{b6}
\end{equation}

\noindent
where $d$ is the spatial dimension (in our case $d=2$).

In order to extend the KW duality symmetry to the continuous field theory
we have need for a "lattice" model definition of the coupling constant $g$,
equivalent to the conventional one exploited in the RG approach. The
renormalization coupling constant $g$ of the $g\Phi^{4}$ theory is closely
related to the fourth derivative of the "Helmholtz free energy", namely
$\partial^{4}F(T,m)/\partial m^{4}$, with respect to the order parameter
$m=\langle \Phi \rangle$. It may be defined as follows
(see \cite{ab8})

\begin{equation}
g(T,h)=-\frac{(\partial^{2}\chi/\partial h^{2})}{\chi^{2}\xi^{d}}+
3\frac{(\partial\chi/\partial h)^{2}}{\chi^{3}\xi^{d}}
\label{b7}
\end{equation}

\noindent
where $\chi$ is the homogeneous magnetic susceptibility

\begin{equation}
\chi=\int d^{2}x G(x)
\label{b8}
\end{equation}

\noindent
It is in fact easy to show that $g(T,h)$ in Eq.(\ref{b7}) is merely the
standard four-spin correlation function taken at zero external momenta.
The renormalized coupling constant $g_{c}$ of the critical theory is defined
by the double limit

\begin{equation}
g_{c}=\lim_{h\rightarrow 0}\lim_{T\rightarrow T_{c}} g(T,h)
\label{b9}
\end{equation}

\noindent
and it is well known that these limits do not commute with each other.
As a result, $g_{c}$ is a path-dependent quantity in the thermodynamic
$(T,h)$ plane \cite{ab8}.

Here we are mainly concerned with the coupling constant on the isochore
line $g(T>T_{c},h=0)$ in the disordered phase and with its critical value

\begin{equation}
g^{*}=\lim_{ T \rightarrow T_{c}^{+}} g(T,h=0)
=-\frac{\partial^{2}\chi /\partial h^{2}}{\chi^{2}\xi^{d}}|_{h=0}
\label{b10}
\end{equation}

\noindent

Here for convenience we have used a simpler notation $g^{*}$ in contrast
to the well-known paper \cite{ab8} where authors made use another symbol
$g^{*}_{+}$.

\section {\normalsize \bf KRAMERS-WANNIER SYMMETRY}
\renewcommand{\theequation}{3.\arabic{equation}}
\setcounter{equation}{0}

The standard KW duality tranformation is known to be as follows
\cite{ab1,ab2,ab3}

\begin{equation}
\sinh(2\tilde{K})=\frac{1}{\sinh(2K)}
\label{b11}
\end{equation}

\noindent
We shall see that it will be more convenient to deal with
a new variable $s=\exp(2K)\tanh(K)$, where $K=J/T$.

It follows from the definition that $s$ transforms as $\tilde{s}=1/s$;
this implies that the correlation length of the 2D Ising model given by
$\xi^{2}=\frac{s}{(1-s)^{2}}$ is a self-dual quantity \cite{ab5c}. Now,
on the one hand, we have the formal relation

\begin{equation}
\xi \frac{ds(g)}{d\xi}=\frac{ds(g)}{dg}\beta(g)
\label{b12}
\end{equation}

\noindent
where $s(g)$ is defined as the inverse function of $g(s)$, i.e. $g(s(g))=g$
and the beta-function is given, as usual, by

\begin{equation}
\xi \frac{dg}{d\xi}=\beta(g)
\label{b13}
\end{equation}

\noindent
On the other hand, it is known from \cite{ab5c}

\begin{equation}
\xi \frac{ds}{d\xi}=\frac{2s(1-s)}{(1+s)}
\label{b14}
\end{equation}

\noindent

From Eq.s (\ref{b12}) - (\ref{b14}), a useful representation of the
beta-function in terms of the $s(g)$ function thus follows

\begin{equation}
\beta(g)=\frac{2s(g)(1-s(g))}{(1+s(g)) \left ( ds(g)/dg \right ) }
\label{b15}
\end{equation}

\noindent
From this representation it follows that $\beta(s(g)_{s(g)=1}=1$
It is reasonable to assume (but it is not obvious!)
that this fixed point is a simple zero.
Then from this equation it would follow that $\omega=\beta'(g^{*})=1$
in agreement with the classical paper \cite{ab16}.

Nevertheless, it should be mentioned that this remarkable result
"contradicts" to conformal field theory results predicting that
for the two-dimensional Ising
model $\omega=\frac{4}{3}$. In the prominent book \cite{ab5b}, being the
Bible for experts in quantum field theory and critical thenomena,
it stands just a question mark instead of $\omega$.
(Another approach one may find in \cite{ab17}) and discussion in \cite{ab5c}).

Let us define the dual coupling constant $\tilde{g}$ and the duality
transformation function $d(g)$ as

\begin{eqnarray}
s(\tilde{g})=\frac{1}{s(g)}; \qquad \qquad \tilde{g} \equiv d(g)
=s^{-1}(\frac{1}{s(g)})
\label{b16}
\end{eqnarray}

\noindent
where $s^{-1}(x)$ stands for the inverse function of $x=s(g)$. It is easy to
check that a further application of the duality map $d(g)$ gives back the
original coupling constant, i.e. $d(d(g))=g$, as it should be. Notice also
that the definition of the duality transformation given by Eq. (\ref{b16})
has a form similiar to the standard KW duality equation, Eq. (\ref{b11}).

Consider now the symmetry properties of $\beta(g)$. We shall see that
the KW duality symmetry property, Eq. (\ref{b11}), results in the
beta-function being covariant under the operation $g\rightarrow d(g)$:

\begin{equation}
\beta(d(g))=d^{\prime}(g)\beta(g)
\label{b17}
\end{equation}

\noindent
To prove it let us evaluate $\beta(d(g))$. Then Eq.(\ref{b15}) yields

\begin{equation}
\beta(d(g))=\frac{2s(\tilde{g})(1-s(\tilde{g}))}{(1+s(\tilde{g}))
\left ( ds(\tilde{g}) / d\tilde{g} \right ) }
\label{b18}
\end{equation}

\noindent
Bearing in mind Eq. (\ref{b16}) one is led to

\begin{equation}
\beta(d(g))=\frac{2s(g)-2}{s(g)(1+s(g))
\left ( ds(\tilde{g}) / d\tilde{g} \right ) }
\label{b19}
\end{equation}

\noindent
The derivative in the r.h.s. of Eq. (\ref{b19}) should be
rewritten in terms of $s(g)$ and $d(g)$. It may be easily done by
applying Eq. (\ref{b16}):

\begin{eqnarray}
\frac{ds(\tilde{g})}{d\tilde{g}}=
\frac{d}{d\tilde{g}}\frac{1}{s(g)}=-\frac{s^{\prime}(g)}{s^{2}(g)}
\frac{1}{d^{\prime}(g)}
\label{b20}
\end{eqnarray}

\noindent
Substituting the r.h.s. of Eq. (\ref{b20}) into Eq. (\ref{b19}) one
obtains the desired symmetry relation, Eq. (\ref{b17}).

Therefore, the self-duality of the model allows us to determine the
fixed point value in another way, namely from the duality equation
$d(g^{*})=g^{*}$.

Making use of a rough approximation for $s(g)$, one gets \cite{ab5c}

\begin{eqnarray}
s(g)\simeq\frac{2}{g}+\frac{24}{g^{2}}\simeq\frac{2}{g}
\frac{1}{1-12/g}=\frac{2}{g-12}
\label{b23}
\end{eqnarray}

\noindent
Combining this Pad\'e-approximant with the definition of $d(g)$,
Eq. (\ref{b16}), one is led to

\begin{eqnarray}
d(g)=4\frac{3g-35}{g-12}
\label{b24}
\end{eqnarray}

\noindent
The fixed point of this function, $d(g^{*})=g^{*}$, is easily seen to
be $g^{*}=14$. The recent numerical and analytical estimates
yield $g^{*}=14.69$ (see \cite{ab5c,ab5d,ab10,ab11} and references therein).

It is worth mentioning that the above-described
approach may be regarded as another method for evaluating $g^{*}$, fully
equivalent to the standard beta-function method.

\section {\normalsize \bf STRONG-WEAK-COUPLING DUALITY}
\renewcommand{\theequation}{4.\arabic{equation}}
\setcounter{equation}{0}

The beta-function of the model under discussion possesses a specific
algebraic property Eq.(\ref{b15}) (KW duality) which allows to construct
the weak-strong-duality transformation $f(g)$ connecting both
the weak-coupling and strong coupling regimes.

Nowadays both the five-loop approximation results (\cite{ab7}) and the strong
coupling expansion for the beta-function \cite{ab5c} are known
rather well. These are given by

\begin{eqnarray}
\beta_{1}(g)&=&2g-2g^{2}+ 1.432347241g^{3}-1.861532885 g^{4}\nonumber\\
&+&3.164776688 g^{5}- 6.520837458 g^{6}+O(g^{7})
\label{b25a}
\end{eqnarray}

\begin{equation}
\beta_{2}(g)=-2g+\frac{12}{\pi}-\frac{9}{\pi^{2}g}+\frac{27}{\pi^{3}g^{2}}
+\frac{81}{8\pi^{4}g^{3}}-\frac{3645}{16\pi^{5}g^{4}}- \frac{15309}{32\pi^{6}g^{5}}
+\frac{2187}{64\pi^{7}g^{6}}+O(g^{-7})
\label{b25b}
\end{equation}

Here indices $1,2$ stand for the weak and strong coupling regimes
respectively. The main goal of this Section is to determine a
dual transformation $f(g)$ such as $f[f(g)]=g$ relating
beta-functions $\beta_{1}(g)$ and $\beta_{2}(g)$.

From Eq.(\ref{b15}) one can easily find the functions $S_{1}(g), S_{2}(g)$
and their inverse functions $G_{1}(s)=S^{-1}_{1}(g), G_{2}(s)=S^{-1}_{2}(g)$
corresponding to two regimes. Simple but cumbersome calculations lead at

\begin{eqnarray}
G_{1}(s)&=&s+s^{2} + 0.3580868104s^{3}-0.1166327797 s^{4}\nonumber\\
&-& 0.1968226859 s^{5}- 0.1299831557 s^{6}+O(s^{7})\nonumber\\
S_{1}(g)&=& g-g^{2} + 1.6419131896 g^{3}-3.09293317 g^{4}\nonumber\\
&+& 6.361881481 g^{5}- 13.78545095 g^{6}+O(g^{7})\nonumber\\
&s&\in[0,1]; \qquad \qquad g\in [0,g^{*}]
\label{b26}
\end{eqnarray}

\newpage

\begin{eqnarray}
G_{2}(s)&=&\frac{3}{4\pi s}+ \frac{9}{2\pi}-\frac{9s}{4\pi}+\frac{18s^{2}}{\pi}-\frac{108s^{3}}{\pi}\nonumber\\
&+&\frac{618s^{4}}{\pi}-\frac{3474s^{5}}{\pi}+\frac{19494s^{6}}{\pi}+O(s^{7})\nonumber\\
S_{2}(g)&=& \frac{2\times3}{8\pi g}+\frac{24\times3^{2}}{(8\pi g)^{2}}
+ \frac{264\times3^{3}}{(8\pi g)^{3}}+\frac{2976\times3^{4}}{(8\pi g)^{4}}\nonumber\\
&+& \frac{35136\times3^{5}}{(8\pi g)^{5}}+ \frac{423680\times3^{6}}{(8\pi g)^{6}}+\frac{5149824\times3^{7}}{(8\pi g)^{7}}+\frac{63275520*3^{8}}{(8\pi g)^{8}}\nonumber\\
&+&O(g^{-9})\nonumber\\
&s&\in [0,1] \qquad \qquad g\in [g^{*},\infty)
\label{b27}
\end{eqnarray}

Having been equipped with these formulas one may easily construct two branches of
the same duality transformation function $f_{12}(g)$ and $f_{21}(g)$ defined
in different domains of $g$. The functions are

\begin{eqnarray}
\frac{1}{f_{21}(g)}&\equiv&\frac{1}{G_{2}(S_{1}(g))}=\frac{4\pi g}{3}-\frac{28 \pi g^{2}}{3}+ 220.5059303 g^{3}\nonumber\\
&-& 1766.8145 g^{4}+ 14816.94007 g^{5}- 127842.5955 g^{6}\nonumber\\
&g&\in [0,g^{*}]; \qquad\qquad f_{21}(g)\in [g^{*},\infty]
\label{b28}
\end{eqnarray}

\begin{eqnarray}
f_{12}(g)&\equiv&G_{1}(S_{2}(g))=\frac{3}{4\pi g} + \frac{63}{16\pi^{2}g^{2}}+ \frac{0.61714739472}{g^{3}}\nonumber\\
&+& \frac{0.9560453953}{g^{4}}+ \frac{1.502156783}{g^{5}}+ \frac{2.368311503}{g^{6}}\nonumber\\
&+&O(g^{7})\nonumber\\
&g&\in [g^{*}, \infty); \qquad\qquad f_{12}(g)\in [0,g^{*}]
\label{b29}
\end{eqnarray}

Functions found above look like inversion, but they are not so trivial.
An interesting nontrivial example of the 2D model disordered
Dirac fermions was discovered in \cite{ab12}. It was shown  that
the beta-function of the (nonintegrable) model under consideration also
exhibits the strong-weak coupling duality such as
$g^{*}\rightarrow \frac{1}{g}$ \cite{ab12}.

It is worth noting that the transformation found is a self-dual indeed

\begin{equation}
f_{12}(f_{21}(g))=f_{21}(f_{12}(g))\equiv g
\label{b30}
\end{equation}

Moreover, by definition weak-strong coupling expansions
of $\beta(g)$ are related to each other in the following way:

\begin{equation}
\beta_{2}(g)=\frac{\beta_{1}(f_{12}(g))}{f^{\prime}_{12}(g)};
\label{b31a}
\end{equation}

\begin{equation}
\beta_{1}(g)=\frac{\beta_{2}(f_{21}(g))}{f^{\prime}_{21}(g)}
\label{b31b}
\end{equation}

These equations are basic results of these article.

\section {\normalsize \bf HIGHER-ORDER TERMS OF THE BETA-FUNCTION}
\renewcommand{\theequation}{5.\arabic{equation}}
\setcounter{equation}{0}

It rather amusing that Eq.(\ref{b29}) looks like a geometric series.
Making use of the Pade method we arrive at

\begin{eqnarray}
f_{12}(g)&\approx& \frac{0.2387324146g^{2}-0.0745907136g+0.0850867165}{g^{3}-1.983571753g^{2}+1.086109562g-0.6919672492}\nonumber\\
&g&\in [g^{*}, \infty); \qquad\qquad f_{12}(g)\in [0,g^{*}]
\label{b32}
\end{eqnarray}

The weak-strong-coupling duality equation and strong-coupling expansion yield
the following numerical values

\begin{eqnarray}
f_{12}(g^{*})-g^{*}&=&0; \qquad\qquad g^{*}=14.38\nonumber\\
\beta_{2}(g^{*})&=&0; \qquad\qquad\qquad g^{*}=14.63
\label{b33}
\end{eqnarray}

being in good agreement with modern estimates \cite{ab13,ab14,ab15}.

Finally, let's consider how one can compute the $\beta(g)$ in the
multiloop approximation via the strong-coupling expansion and the S-duality
function. In order to find that one should exploit Eq.(\ref{b31a}),Eq.(\ref{b25b})
and the approximate expression for $f_{12}(g)$ given by Eq.(\ref{b32}).

After some tedious but routine calculations we arrive to some
polynomial of 7th degree for $\beta_{1}(g)$.

\begin{eqnarray}
\beta_{1}(g)&=&2g-2g^{2}+ 1.432347241g^{3}-1.861532885 g^{4}\nonumber\\
&+&3.164776688 g^{5}- 6.520837458 g^{6}- 331.454743 g^{7}
\label{b34}
\end{eqnarray}

It is easily seen that the first 6 terms excepting for the 7th one
are the exact perturbation expansion for $\beta_{1}(g)$ \cite{ab7}.
It would be tempting but wrong to regard Eq.(\ref{b34})
as a $\beta(g)$-function in the 7th loop approximation.
In fact, the function in Eq.(\ref{b32}) is approximate,
so that we have to estimate an accuracy of our calculations.

Let's illustrate our treatment by using a simple example.
Suppose, that a difference between the "exact" duality function
$f_{12}^{exact}(g)$ and the approximate one given by Eq.(\ref{b32}) reads

\begin{equation}
f_{12}^{exact}(g)= \frac{0.2387324146g^{2}-0.0745907136g+0.0850867165}{g^{3}-1.983571753g^{2}+1.086109562g-0.6919672492}+b/g^{7}
\label{b35}
\end{equation}

with $b$ being an arbitrary unknown parameter. The straightforward
calculation shows
that a "new"  7th loop contribution computed by making use
of the Eq.(\ref{b35}) depends on the fitting parameter $b$ and
differs vastly from the previous one, it being

\begin{eqnarray}
\beta_{1}(g)&=&2g-2g^{2}+ 1.432347241g^{3}-1.861532885 g^{4}\nonumber\\
&+&3.164776688 g^{5}- 6.520837458 g^{6}+(- 331.454743+271519.803807b)g^{7}
\label{b36}
\end{eqnarray}

Assuming for instance that $b=10^{-4}$, so that the difference
between exact dual function $f_{12}^{exact}(g)$ and a trial one
is numerically small indeed, one obtains that a change of the sixth
loop coefficient is of order 1.

Thus, we see that the approach suggested above provides a regular scheme for
computing higher-order corrections to the $\beta(g)$-function on
the basis of lattice high-order expansions.
In other words, one obtaines a tempting possibility to compute
(approximately) multiloop Feynman diagrams on the basis of Eq.(\ref{b31b})
and of high-temperature expansions \cite{ab5c}. The serious drawback
of that scheme is that it being unstable from a mathematical point of view.

\section{\normalsize \bf CONCLUDING REMARKS}
\renewcommand{\theequation}{6.\arabic{equation}}
\setcounter{equation}{0}

We have shown that the $\beta$-function of the $2D$  $g\Phi^{4}$ theory
does have two types of dual symmetries, these being the Kramers-Wannier
symmetry and the weak-strong-coupling symmetry or S-duality.
These important symmetries were shown to have a thorough
mathematical background, it being highly nontrivial algebraic structure
Eq.(\ref{b15}) of $\beta(g)$.

Our proof of the KW symmetry is based on the properties of $g(s),s(g)$
defined only for $ 1\leq s < \infty; g^{*}\leq g< \infty$ and therefore
doesn't cover the weak-coupling region, $0\leq g \leq g^{*}$. So that the
statement is that the beta-function $\beta(g)$ possesses the KW symmetry
only in the strong-coupling region.

In contrast to widely held views, the KW symmetry imposes only mild
restrictions on $\beta(g)$. It means that this symmetry property fixes
only even derivatives of the beta-function $\beta^{(2k)}(g^{*}) (k=0,1,...)$
at the fixed point, leaving the odd derivatives free, in particular, the
critical exponent $\omega$, responsible for corrections to scaling.

We have established the existence of the nontrivial weak-strong-coupling
dual function $f(g)$ connecting two domains of both weak coupling and
strong coupling. Given both perturbative RG calculations
and lattice high-temperature expansions that S-function
$f(g)$ can be approximately computed. We also explicitly computed
high-order terms of $\beta(g)$.
In summary, the value of the above developed approach is considerably deduced
by its being strongly unstable.

\section{\normalsize \bf ACKNOWLEDGEMENTS}

The author is most grateful to
Istituto Nazionale di Fisica Nucleare, Struttura di Pavia
and the Max-Planck Institute in Stuttgart
for kind hospitality and the use of its facilities.
He has much benefitted from numerous helpful discussions with
G.Jug, G.Khaliullin, A.I.Sokolov, E.V.Orlov, and K.B.Varnashev.

I also thank to the Russion Foundation for Basic Research
(grant N 01-02-17794) for financial support.

\end{document}